\def\be{\begin{equation}}
\def\ee{\end{equation}}
\def\bea{\begin{eqnarray}}
\def\eea{\end{eqnarray}}
\def\bse{\begin{subequations}}
\def\ese{\end{subequations}}

\documentclass[twocolumn,prb,amsmath,amssymb,eqsecnum,preprintnumbers]{revtex4}
\usepackage{graphicx} % Include figure files
\usepackage{dcolumn} % Align table columns on decimal point
\usepackage{bm} % bold math
\usepackage{amsfonts}

\draft
\begin{document}
%\preprint{KITP Preprint NSF-KITP-05-xx}
%\preprint{Phys. Rev. B {\bf 74}, 024409 (2006)}
\title{A compilation of metallic systems that show a quantum ferromagnetic transition}
\author{D. Belitz$^{1,2}$ and T. R. Kirkpatrick$^3$}
\affiliation{$^{1}$ Department of Physics and Institute of Theoretical Science,
           University of Oregon, Eugene, OR 97403, USA\\
$^{2}$ Materials Science Institute, University of Oregon, Eugene, OR 97403, USA\\
$^{3}$ Institute for Physical Science and Technology,and Department of 
                             Physics, University of Maryland, College Park, MD 20742, USA\\
}
\date{\today}

\begin{abstract}
We provide a compilation of metallic systems in which a low-temperature ferromagnetic
or similar transition is observed. Our objective is to demonstrate the universal first-order
nature of such transitions in clean systems in two or three spatial dimensions. Please 
contact the authors with information about omissions, corrections, or any other information.
\end{abstract}
%\pacs{}
\maketitle
%\narrowtext
%\tableofcontents
%\begin{widetext}
Quantum phase transitions are phenomena of great interest.\cite{Hertz_1976,
Sachdev_1999} Perhaps the most obvious quantum phase transition, and the first one
considered historically, is the transition from a paramagnetic metal to a ferromagnetic
metal at zero temperature ($T=0$) as a function of some non-thermal control parameter. Stoner's
theory of itinerant ferromagnetism\cite{Stoner_1938} describes both the thermal
transition and the static properties of the quantum transition in a mean-field approximation.
It predicts a second-order or continuous transition with standard Landau or
mean-field static critical exponents. For the thermal or classical transition this constitutes an
approximation for spatial dimensions $d\leq 4$. In the physical dimensions $d=3$ or lower,
fluctuations of the magnetization order parameter, which are neglected in Stoner theory, 
lead to deviations from the mean-field critical behavior that require the renormalization 
group (RG) for a theoretical understanding.\cite{Wilson_Kogut_1974} 
In a seminal paper, Hertz\cite{Hertz_1976} derived a Landau-Ginzburg-Wilson (LGW) functional
for the ferromagnetic transition from a model of itinerant electrons that interact via
a point-like potential in the particle-hole spin-triplet channel. Hertz analyzed this
dynamical LGW functional by means of RG methods.
He concluded that, at $T=0$, Stoner theory is exact as far as the static critical 
behavior is concerned, i.e., the transition
is of second order with mean-field static critical exponents, and a dynamical critical
exponent $z=3$, for all $d>1$. This is because the coupling
of the statics to the dynamics makes the system effectively behave as if it were in a higher
spatial dimension, given by $D = d+z$.

\begin{figure}[b]
\vskip -0mm
\includegraphics[width=8.0cm]{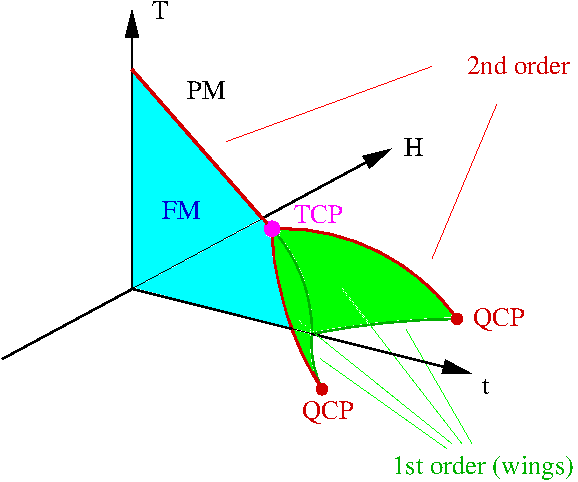}
\caption{Generic phase diagram of a metallic ferromagnet in the space spanned 
               by temperature ($T$), magnetic field ($H$), and the control parameter ($t$). Shown 
               are the ferromagnetic (FM) and paramagnetic (PM) phases, 
               lines of second-order transitions, surfaces of first-order transitions 
               (``tricritical wings''), the tricritical point (TCP), and the two quantum 
               critical points (QCP).}
\label{fig:1}
\end{figure}

It became clear in the late 1990s that this theoretical picture is not correct. It was shown
that particle-hole excitations about the Fermi surface, which exists in all metals in
dimensions $d>1$, couple to the magnetization and invalidate Hertz's 
conclusions.\cite{Belitz_Kirkpatrick_Vojta_1999, Belitz_Kirkpatrick_Vojta_2005} As a result, the quantum ferromagnetic
transition was predicted to be generically of first order in clean metallic ferromagnets in
$d>1$. Physically, the mechanism that drives the transition first order is very similar to
the fluctuation-induced first-order transition that was predicted earlier for the
classical transition in superconductors and smectic liquid 
crystals,\cite{Halperin_Lubensky_Ma_1974} and to the spontaneous mass-generation
mechanism known as the Coleman-Weinberg mechanism in particle 
physics.\cite{Coleman_Weinberg_1973} In all of these cases a generic soft mode
(the photon in the cases of superconductors and scalar electrodynamics; the nematic
Goldstone mode in the case of liquid crystals) that is distinct from the order parameter
fluctuations couples to the latter and qualitatively changes the nature of the phase
transition. For the quantum ferromagnetic transition in metals, the resulting prediction is 
the generic phase diagram shown in Fig.\ \ref{fig:1}.
%%%%%%%%%%%%%%%%%%%
%%%%%%%%%%%%%%%%%%%
At zero temperature ($T=0$), there is a first-order transition triggered by a non-thermal
control parameter $t$. A nonzero temperature gives the generic particle-hole excitations
a mass, and as a result the mechanism driving the first-order transition becomes weaker
with increasing temperature. This leads to a tricritical point (TCP) at some temperature 
$T_{\text{tc}} > 0$, and for $T > T_{\text{tc}}$ the transition is generally of second order
with classical critical exponents. Upon the application of an external field conjugate to 
the order parameter, i.e., a homogeneous magnetic field $H$ in the case of a ferromagnet,
surfaces of first-order transitions called tricritical wings emerge from the tricritical
point. This is true for any classical phase diagram that contains a tricritical
point,\cite{Griffiths_1970} and it holds for quantum ferromagnets as 
well.\cite{Belitz_Kirkpatrick_Rollbuehler_2005} These tricritical wings are bounded by
lines of second-order transitions and end in a pair of quantum critical points (QCPs)
in the $T=0$ plane. The critical behavior at these QCPs can be determined exactly and
is a slight modification of the critical behavior predicted by Hertz for the quantum phase
transition at $H=0$ that is pre-empted by the first-order transition.\cite{Belitz_Kirkpatrick_Rollbuehler_2005} 

This general picture is theoretically predicted to apply to all transitions from a metallic
paramagnetic phase to a metallic ferromagnetic one in
dimensions $d>1$, irrespective of whether the magnetism is caused by the conduction
electrons (``itinerant ferromagnets'') of by electrons in a different band, and irrespective
of the isotropy or lack thereof of the magnetization. That is, it applies to
easy-axis (Ising) and easy-plane (XY) magnets as well as to isotropic (Heisenberg) magnets.
It also applies to ferrimagnets and canted ferromagnets, and more generally to any metallic 
system that has a nonvanishing homogeneous magnetization,\cite{Kirkpatrick_Belitz_2012b}
but the most extensive experimental information is available for ferromagnets. There are
only two ways to avoid these conclusions: (1) In 1-d or quasi-1-d systems there is no
Fermi surface, and hence no particle-hole excitations, and the soft-mode mechanism
is not operative. (2) In the presence of sufficiently strong quenched disorder the nature
of the particle-hole excitations changes from ballistic to diffusive, and the nature of
the coupling to the magnetization changes as well. This can lead to a second-order quantum
phase transition with non-mean-field critical exponents that still can be determined
exactly.\cite{Kirkpatrick_Belitz_1996, Belitz_et_al_2001a, Belitz_et_al_2001b,
Belitz_Kirkpatrick_Vojta_2005} Disorder
may also have stronger effects, leading to Griffiths phases and smeared transitions,
see Ref.\ \onlinecite{Vojta_2010}.

Experimentally, the picture summarized above is confirmed with remarkable uniformity.
To the authors's knowledge, all metallic ferromagnets that do not fall into one of the
two exceptional classes mentioned above, show a first-order transition if the transition
temperature is sufficiently low, or can be driven sufficiently low by a non-thermal
control parameter, such as pressure, or composition. This is especially
remarkable if compared with the case of classical liquid crystals, where the observed
transition is usually of second order, and only recently have examples of weakly
first-order transitions been found.\cite{Yethiraj_Mukhopadhyay_Bechhoefer_2002}
The reason why the theory is so much more successful in the quantum case is not
entirely understood, but it is likely related to the fact that order-parameter fluctuations,
which can invalidate the fluctuation-induced first-order mechanism, are strongly
suppressed in the quantum case for the same reasons that lead to a mean-field
critical behavior in Hertz's theory.\cite{Kirkpatrick_Belitz_2012b} 

The purpose of this informal communication is to demonstrate this remarkable agreement
between theory and experiment by compiling a list of metallic systems in which a quantum 
ferromagnetic transition has been observed. 
%The table lists all examples the authors are
%aware of; please contact us if you notice omissions, or if you have corrections or other
%comments. 
The systems are listed roughly in order of completeness of the experimental
information available. All but three of the systems listed display a confirmed or suspected
first-order transition. The three expections are, URu$_{2-x}$Re$_x$Si$_2$, which is
strongly disordered, YbNi$_4$P$_2$, which is quasi-1-d, and Ni$_x$Pd$_{1-x}$, where
the lowest transition temperature achieved is 7K, which may be above the tricritical
point, if one exists. All other examples are consistent with the phase diagram shown in
Fig.\ \ref{fig:1}. In some cases (e.g., UCoGe) the transition
is first order at the highest, or only, temperature observed, so the tricritical point is not
accessible. In all cases where a tricritical point is accessible and the behavior in a 
magnetic field has been studied, tricritical wings have been observed. One of the best
studied materials, MnSi, is actually a helimagnet,\cite{Ishikawa_et_al_1976} but the 
helical wavelength ($\approx 200\AA$) is so long compared to the atomic length
scale that the system is well approximated as a ferromagnet.\cite{Pfleiderer_et_al_1997}

We conclude with a few general remarks. First, there also are cases of transitions from a
metallic ferromagnet to some insulating phase. Examples include, 
FeSi$_{1-x}$Ge$_x$,\cite{Yeo_et_al_2003} and RE$_{0.55}$Sr$_{0.45}$Mn$_3$, with RE a 
rare earth or a combination of rare earths.\cite{Demko_et_al_2008} In these cases the
theoretical situation is more complicated, and we do not include them in our discussion.
Second, in any given material a first-order transition may occur for reasons other than the
coupling to particle-hole excitations. This is likely the case in systems that have a
tricritical point at a relatively high temperature, such as various 
manganites, see, e.g., Ref.\ \onlinecite{Kim_et_al_2002}. Finally, we mention that some
of the materials listed in the table have gotten a lot of attention for properties other than,
although possibly related to, the ferromagnetic transition. Examples are the coexistence
of superconductivity and ferromagnetism observed in UGe$_2$,\cite{Saxena_et_al_2000}
URhGe,\cite{Aoki_et_al_2001} and UCoGe,\cite{Huy_et_al_2007} or the non-Fermi-liquid
phase and the A-phase in MnSi.\cite{Pfleiderer_Julian_Lonzarich_2001, Muehlbauer_et_al_2009}
As a result, there is a large body of literature on some of these materials; we quote
only papers that are directly relevant to properties reflected by the entries in the table.

\acknowledgments
We gratefully acknowledge discussions and correspondence with Greg Stewart, Jeff Lynn, Nick Butch,
and Christoph Geibel. 
This work was supported by the National Science Foundation under Grant Nos. 
DMR-09-29966, and DMR-09-01907.
\onecolumngrid

%\vfill\eject
%%%%%%%%%%%%%%%%%%%%%%%%%%%%%%%%%%%%%%%%%%%%%%%%%%%%%%%%%%%%%%%%%%%%%
\begin{table*}[!]
\caption{Systems with low-$T$ ferromagnetic transitions and their properties. $T_{\text{c}} =$
 Curie temperature, $T_{\text{tc}} =$ tricritical temperature. $\rho_0 =$ 
 residual resistivity. FM = ferromagnet, SC = superconductor. N/A = not applicable; n.a. = not available.}
\smallskip
\begin{ruledtabular}
\begin{tabular}{ccccccccc}
%\\
System \footnote{References in this column refer to reviews, if any exist. Most references are to be
                            understood  as ``This reference and references therein".}
   & Order of 
      & $T_{\text{c}}$/K \footnote{A single value of $T_{\text{c}}$, for the default value of the tuning parameter (ambient pressure, zero field) is given where a tricritical temperature has also been measured. A range of $T_{\text{c}}$, with a corresponding range of the control parameter, is given in all other cases.} 
         & magnetic 
            & tuning  
               & $T_{\text{tc}}$/K
                  & wings 
                     & Disorder  
                        & Comments \\

   & Transition\footnote{At the lowest temperature achieved.}  
      &                            
         & moment/$\mu_{\text{B}}$\footnote{Per formula unit unless otherwise noted.}
            &  parameter 
               &                              
                  & observed  
                     & ($\rho_0/\mu\Omega$cm)\footnote{For the highest-quality samples.} 
                        & \\
%\\
\hline\\
%%%%%%%%%%%%%%%%%%%%%%%%%%%%%%%
MnSi$\ $\cite{Pfleiderer_2007}
   & 1st$\ $\cite{Pfleiderer_et_al_1997} 
      & $29.5\ $\cite{Ishikawa_et_al_1985} 
         & 0.4$\ $\cite{Ishikawa_et_al_1985} 
            & hydrostatic 
               & $\approx 10\ $\cite{Pfleiderer_et_al_1997}  
                  & yes$\ $\cite{Pfleiderer_Julian_Lonzarich_2001}
                     & $0.33\ $\cite{Pfleiderer_Julian_Lonzarich_2001}
                        & weak helimagnet$\ $\cite{Ishikawa_et_al_1976}\\
   &
      & 
         &
            & pressure$\ $\cite{Pfleiderer_et_al_1997}  
               & 
                  & 
                     & 
                        & exotic phases$\ $\cite{Pfleiderer_Julian_Lonzarich_2001, Muehlbauer_et_al_2009} \\
\\
%%%%%%%%%%%%%%%%%%%%%%%%%%%%%%%%%
ZrZn$_2$$\ $\cite{Pfleiderer_2007} 
   & 1st$\ $\cite{Uhlarz_Pfleiderer_Hayden_2004}
      & 28.5$\ $\cite{Uhlarz_Pfleiderer_Hayden_2004} 
         & 0.17$\ $\cite{Uhlarz_Pfleiderer_Hayden_2004}  
            & hydrostatic
               & $\approx 5$$\ $\cite{Uhlarz_Pfleiderer_Hayden_2004} 
                  & yes$\ $\cite{Uhlarz_Pfleiderer_Hayden_2004}  
                     & $\geq 0.31$$\ $\cite{Sutherland_et_al_2012}  
                        & confusing history, \\

   & 
      & 
         & 
            & pressure$\ $\cite{Uhlarz_Pfleiderer_Hayden_2004}  
               & 
                  & 
                     & 
                        & see Ref.\ \onlinecite{Pfleiderer_2007}  \\
\\
%%%%%%%%%%%%%%%%%%%%%%%%%%%%%%%%%%
Sr$_3$Ru$_2$O$_7$
   & 1st \footnote{Phase diagram not mapped out completely; the most detailed measurements show
                               the tips of the wings. See Ref.\ \onlinecite{Wu_et_al_2011}.}
      & 0 \footnote{Paramagetic at ambient pressure. Hydrostatic pressure drives the
                                              system away from FM, uniaxial stress drives it towards FM. See
                                              Ref.\ \onlinecite{Wu_et_al_2011} and references therein, especially
                                              Ref.\ \onlinecite{Ikeda_et_al_2000}.}  
         & 0$^{\ g}$
            & pressure$^{\ g}$
               & n.a.
                  & yes$\ $\cite{Wu_et_al_2011}  
                     & $< 0.5$$\ $\cite{Wu_et_al_2011}  
                        & foliated wing tips,\\

   & 
      & 
         & 
            &   
               & 
                  & 
                     & 
                        & nematic phase$\ $\cite{Wu_et_al_2011} \\
\\
%%%%%%%%%%%%%%%%%%%%%%%%%%%%%%%%%%%%
UGe$_2$$\ $\cite{Aoki_et_al_2011} 
   & 1st$\ $\cite{Huxley_et_al_2001} 
      & 52$\ $\cite{Kotegawa_et_al_2011} 
         & 1.5$\ $\cite{Kotegawa_et_al_2011}  
            & hydrostatic 
               & 24$\ $\cite{Taufour_et_al_2010} 
                  & yes$\ $\cite{Taufour_et_al_2010, Kotegawa_et_al_2011} 
                     & 0.2$\ $\cite{Saxena_et_al_2000} 
                        & easy-axis FM \\

   &       
      & 
         & 
            & pressure$\ $\cite{Saxena_et_al_2000, Kotegawa_et_al_2011}  
               & 
                  & 
                     & 
                        & coexisting FM+SC$\ $\cite{Saxena_et_al_2000}\\
\\
%%%%%%%%%%%%%%%%%%%%%%%%%%%%%%%%%%%%
URhGe$\ $\cite{Aoki_et_al_2011} 
   & 1st$\ $\cite{ Huxley_et_al_2007}  
      & 9.5$\ $\cite{Aoki_et_al_2001} 
         & 0.42$\ $\cite{Aoki_et_al_2001}  
            &  transverse 
               & $\approx 1$$\ $\cite{ Huxley_et_al_2007} 
                  & yes$\ $\cite{ Huxley_et_al_2007} 
                     & 8$\ $\cite{Miyake_Aoki_Flouquet_2009} 
                        & easy-plane FM\\

   & 
      & 
         & 
            & $B$-field$\ $\cite{Levy_et_al_2005, Huxley_et_al_2007}
               & 
                  & 
                     & 
                        & coexisting FM+SC$\ $\cite{Aoki_et_al_2001} \\
\\
%%%%%%%%%%%%%%%%%%%%%%%%%%%%%%%%%%%%%%%
UCoGe$\ $ \cite{Aoki_et_al_2011}
   & 1st$\ $\cite{ Hattori_et_al_2009} 
      & 2.5$\ $\cite{ Hattori_et_al_2009}  
         & 0.03$\ $\cite{ Huy_et_al_2007}  
            & none 
               & $>2.5$? \footnote{1st  order transition with no tuning parameter; TCP not accessible.}
                  & no 
                     & 12$\ $\cite{ Huy_et_al_2007}   
                        & coexisting FM+SC$\ $\cite{Huy_et_al_2007} \\
\\
%%%%%%%%%%%%%%%%%%%%%%%%%%%%%%%%%%%%%%%%
CoS$_2$ 
   & 1st$\ $\cite{Sidorov_et_al_2011}  
      & 122$\ $\cite{Sidorov_et_al_2011}  
         & 0.84$\ $\cite{Sidorov_et_al_2011}  
            & hydrostatic 
               & $\approx 120$$\ $\cite{Sidorov_et_al_2011}  
                  & no 
                     & 0.7$\ $\cite{Sidorov_et_al_2011}   
                        & rather high $T_{\text{c}}$\\

   & 
      & 
         & 
            & pressure$\ $\cite{Sidorov_et_al_2011}  \\
\\
%%%%%%%%%%%%%%%%%%%%%%%%%%%%%%%%%%%%%%%%%%
La$_{1-x}$Ce$_x$In$_2$
   & 1st$\ $\cite{Rojas_et_al_2011}  
      & 22 -- 19.5$\ $\cite{Rojas_et_al_2011}  \footnote{For $x = 1.0$ -- $0.9$} 
         & n.a.
            & composition$\ $\cite{Rojas_et_al_2011} 
               & $>22$? \footnote{1st order for $x=1$, TCP not accessible.}
                  & no 
                     & n.a. 
                        & third phase between\\

   & 
     &  
        & 
           & 
              & 
                 & 
                    & 
                       & FM and PM?$\ $\cite{Rojas_et_al_2011}\\
\\
%%%%%%%%%%%%%%%%%%%%%%%%%%%%%%%%%%%%%%%%%%
Ni$_3$Al$\ $\cite{Pfleiderer_2007} 
   & (1st) \footnote{Suspected 1st order transition near $p=80$kbar, Refs.\ 
                               \onlinecite{Niklowitz_et_al_2005, Pfleiderer_2007}.}
      & 41 -- 15 \footnote{For pressures $p=0$ -- $60$ kbar, Ref.\ \onlinecite{Niklowitz_et_al_2005}.} 
         & 0.075 \footnote{Per Ni at $p=0$, Ref.\ \onlinecite{Niklowitz_et_al_2005}} 
            & hydrostatic 
               & n.a. 
                  & no 
                     & 0.84$\ $\cite{Steiner_et_al_2003} 
                        & order of transition\\

   & 
      &  
         & 
            & pressure$\ $\cite{Niklowitz_et_al_2005} 
               & 
                  & 
                     & 
                        &  uncertain \\
\\
%%%%%%%%%%%%%%%%%%%%%%%%%%%%%%%%%%%%%%%%%%
YbIr$_2$Si$_2$ \footnote{YbRh$_2$Si$_2$ belongs to the same family, but has an AFM phase
                                          between the FM and the PM.\cite{Yuan_et_al_2006}.}
   & 1st$\ $\cite{Yuan_et_al_2006} 
      & 1.3 -- 2.3 \footnote{For pressures $p \approx 8$ -- $10$GPa.}  
         & n.a. 
            & hydrostatic 
               & n.a. 
                  & no 
                     & $\approx 22$ \footnote{For a magnetic sample at pressures $p \approx 8$ -- $10$GPa.
                                                               Samples with
                                                               $\rho_0$ as low as 0.3$\mu\Omega$cm have been
                                                               prepared.\cite{Yuan_et_al_2006}}  
                        & FM nature of ordered \\

   & 
      & 
         & 
            & pressure$\ $\cite{Yuan_et_al_2006} 
               & 
                  & 
                     & 
                        & phase suspected$\ $\cite{Yuan_et_al_2006} \\
\\
%%%%%%%%%%%%%%%%%%%%%%%%%%%%%%%%%%%%%%%%%%
YbCu$_2$Si$_2$$\ ^{n}$
   & n.a. 
      & 4 -- 6$\ $\cite{Winkelmann_et_al_1999} \footnote{For pressures $p \approx 10$ -- $20$GPa.}
         & n.a.
             & hydrostatic
                & n.a.
                   & no
                      & n.a.
                         &  nature of magnetic\\

   &
      &
         &
            & pressure$\ $\cite{Winkelmann_et_al_1999}
                &
                   &
                      &
                         &  order unclear\\
\\
%%%%%%%%%%%%%%%%%%%%%%%%%%%%%%%%%%%%%%%%%%%%
%La$_x$Ce$_{1-x}$Ru$_2$Si$_2$
%   & n.a. 
%      & 6 -- 1$\ $\cite{Flouquet_et_al_1995}  \footnote{For $x=0.20$ -- $0.075$} 
%         & 1.2 -- 0.25 $^p$ 
%            & composition 
%               & n.a. 
%                  & no 
%                     & 4$\ $\cite{Flouquet_et_al_1995} 
%                        & order of transition\\
%
%   & 
%      & 
%         & 
%            & 
%               & 
%                  & 
%                     & 
%                        & not known \\ 
%\\
\hline
\\
%%%%%%%%%%%%%%%%%%%%%%%%%%%%%%%%%%%%%%%%%%%%
URu$_{2-x}$Re$_x$Si$_2$ 
   & 2nd$\ $\cite{Bauer_et_al_2005, Butch_Maple_2009} 
      & 25 -- 2 \footnote{for $x=0.6$ -- $0.2$, Ref.\ \onlinecite{Butch_Maple_2009}.} 
         & 0.4 -- 0.03$\ $\cite{Butch_Maple_2009}
            & composition$\ $\cite{Bauer_et_al_2005} 
               & N/A 
                  & N/A 
                     & $\approx 100$ \footnote{For $x=0.1$, Ref.\ \onlinecite{Butch_Maple_2010}.} 
                        & strongly disordered\\
\\
%%%%%%%%%%%%%%%%%%%%%%%%%%%%%%%%%%%%%%%%%%
Ni$_x$Pd$_{1-x}$ 
   & 2nd$\ $\cite{Nicklas_et_al_1999}
      & 600 -- 7 \footnote{for $x = 1$ -- $0.027$, Ref.\ \onlinecite{Nicklas_et_al_1999}} 
         & n.a. & composition$\ $\cite{Nicklas_et_al_1999} 
            & N/A 
               & N/A 
                  & n.a. 
                     & disordered, lowest \\
& & & & & & & & $T_{\text c}$ rather high \\
\\
%%%%%%%%%%%%%%%%%%%%%%%%%%%%%%%%%%%%%%%%%%
YbNi$_4$P$_2$ 
   & 2nd$\ $\cite{Krellner_et_al_2011}  
      & 0.17$\ $\cite{Krellner_et_al_2011}  
         & $\approx 0.05$$\ $\cite{Krellner_et_al_2011}  
            & none 
               & N/A 
                  & N/A 
                     & 2.6$\ $\cite{Krellner_et_al_2011}  
                        & quasi-1d, disordered\\
\\
\end{tabular}
\end{ruledtabular}
\end{table*}
\vfill\eject
%%%%%%%%%%%%%%%%%%%%%%%%%%%%%%%%%%%%%%%%%%%%%%%%%%%%%%%%%%%%%%%%%%%%%%%
\twocolumngrid
%\bibliography{table}

\end{document}